\documentclass[journal,onecolumn]{IEEEtran}
\usepackage{epsfig,color,amsmath}
\usepackage{amsthm}
\usepackage{wrapfig}
\usepackage{bm}
\usepackage{epstopdf}
\usepackage{lipsum} 
\usepackage{makecell}
\usepackage{amssymb}
\usepackage{url}
\usepackage{mathtools}
\usepackage{enumitem}
\usepackage{multirow}
\usepackage{makecell}
\usepackage{amsmath}
\usepackage{stackrel}
\setlist[itemize]{leftmargin=*}

\makeatother

\newcommand{\m}{\boldsymbol}
\allowdisplaybreaks[4]
\usepackage[colorlinks=true,citecolor=red,linkcolor=black,urlcolor  = blue,anchorcolor = blue]{hyperref}

\newcommand{\mc}[1]{\mathcal{#1}}
\newcommand{\mbb}[1]{\mathbb{#1}}

\usepackage{graphicx}
\usepackage{ifpdf}
\usepackage{epstopdf}
\usepackage{ifpdf}
\ifpdf
\DeclareGraphicsExtensions{.eps}
\else
\DeclareGraphicsExtensions{.eps}
\fi

\usepackage{pgf}

\usepackage{stackengine} 
\stackMath
\usepackage[flushleft]{threeparttable}
\usepackage{makecell}
\usepackage{tabularx}
\usepackage{multicol}
\usepackage{mathtools}

\usepackage[normalem]{ulem}
\usepackage{mathrsfs}
\DeclareMathAlphabet\mathbfcal{OMS}{cmsy}{b}{n}
\DeclarePairedDelimiter\norm{\lVert}{\rVert}%

\usepackage{stackengine}

\usepackage{etoolbox}
\usepackage[procnumbered,linesnumbered,lined,boxed,commentsnumbered,ruled,longend]{algorithm2e}


\makeatletter
\AtBeginEnvironment{procedure}{\let\c@algocf\c@procedure}
\makeatother

\usepackage{mathtools}

\usepackage{tikz}
\usetikzlibrary{patterns,tikzmark}
\usetikzlibrary{matrix,decorations.pathreplacing,calc}
\usepackage{hf-tikz}
\usepackage{amsthm}
\usepackage[export]{adjustbox}

\usepackage[utf8]{inputenc}
\usepackage[T1]{fontenc}
\usepackage{lmodern}
\usepackage[figurename=Fig.,labelfont=bf,labelsep=period]{caption}
\usepackage[colorlinks=true,citecolor=red,linkcolor=black]{hyperref}
\usepackage{times}

\title{{Data-Driven Identification of Dynamic Quality Models in Drinking Water Networks}}
\author{Shen Wan$\text{g}^{\dagger}$, Ankush Chakrabart$\text{y}^\ddagger$, and Ahmad F. Tah$\text{a}^{\diamond,\ast}$
\thanks{$^\dagger$Lecturer, School of Cyberspace Security, Beijing University of Posts and Telecommunications, Beijing 100876, China. Email: shen.wang@bupt.edu.cn}
\thanks{$^\ddagger$Research Scientist, Department of Control and Dynamical Systems, Mitsubishi Electric Research Laboratories. Email: chakrabarty@merl.com}
\thanks{$^\diamond$Associate Professor, Department of Civil and Environmental Engineering at Vanderbilt University, Nashville, TN 37235, USA. Email: ahmad.taha@vanderbilt.edu}
\thanks{$^\ast$Corresponding author. }
\thanks{This work is partially supported by the National Natural Science Foundation of China under Grant  62203062 and the National Science Foundation under Grant 2151392.}
}
\makeatother
\begin{document}
	
	\maketitle
	
	\setlength{\abovedisplayskip}{3.5pt}
	\setlength{\belowdisplayskip}{3.5pt}
	\setlength{\abovedisplayshortskip}{3.1pt}
	\setlength{\belowdisplayshortskip}{3.1pt}
	
	\newdimen\origiwspc%
	\newdimen\origiwstr%
	\origiwspc=\fontdimen2\font
	\origiwstr=\fontdimen3\font
	
	\fontdimen2\font=0.64ex
\begin{abstract}
 Traditional control and monitoring of water quality in drinking water distribution networks (WDN) rely on mostly model- or toolbox-driven approaches, where the network topology and parameters are assumed to be known. In contrast, system identification (SysID) algorithms for generic dynamic system models seek to approximate such models using only input-output data without relying on network parameters. The objective of this paper is to investigate SysID algorithms for water quality model approximation. This research problem is challenging due to \textit{(i)} complex water quality and reaction dynamics and \textit{(ii)} the mismatch between the requirements of SysID algorithms and the properties of water quality dynamics. In this paper, we present the first attempt to identify water quality models in WDNs using only input-output experimental data and classical SysID methods without knowing any WDN parameters. Properties of water quality models are introduced, the ensuing challenges caused by these properties when identifying water quality models are discussed, and remedial solutions are given. Through case studies, we demonstrate the applicability of SysID algorithms, show the corresponding performance in terms of accuracy and computational time, and explore the possible factors impacting water quality model identification.
\end{abstract}


	\section*{List of Acronyms}
\begin{table}[h]
\normalsize
\begin{tabular}{l l}
WDN & Water Distribution Network\\
SysID &  System Identification \\
SIM & Subspace Identification Method\\
CVA & Canonical Variate Analysis\\
MOESP & Multivariable Output-error State Space\\
N4SID & \makecell{Numerical subspace state space \\System Identification}\\ 
ERA & Eigensystem Realization Algorithm \\
OKID & Observer/Kalman filter Identification\\
SVD & Singular Value Decomposition \\
WQMI & Water Quality Model Identification
\end{tabular}
\end{table}
	\section{Introduction and Paper Contributions}~\label{sec:intro}
Water quality modeling describes the spatiotemporal evolution of disinfectant concentration (e.g., chlorine) in various network elements in drinking water distribution networks (WDN). However, the water quality model can still be uncertain due to the complicated reaction dynamics that are severely affected by various factors including topology, parameters, uncertainty (e.g., sensor measurement noise, demand estimation errors), and leakage~\cite{9394797}.

\begin{figure}[t]
	\centering
	\includegraphics[width=0.45\linewidth]{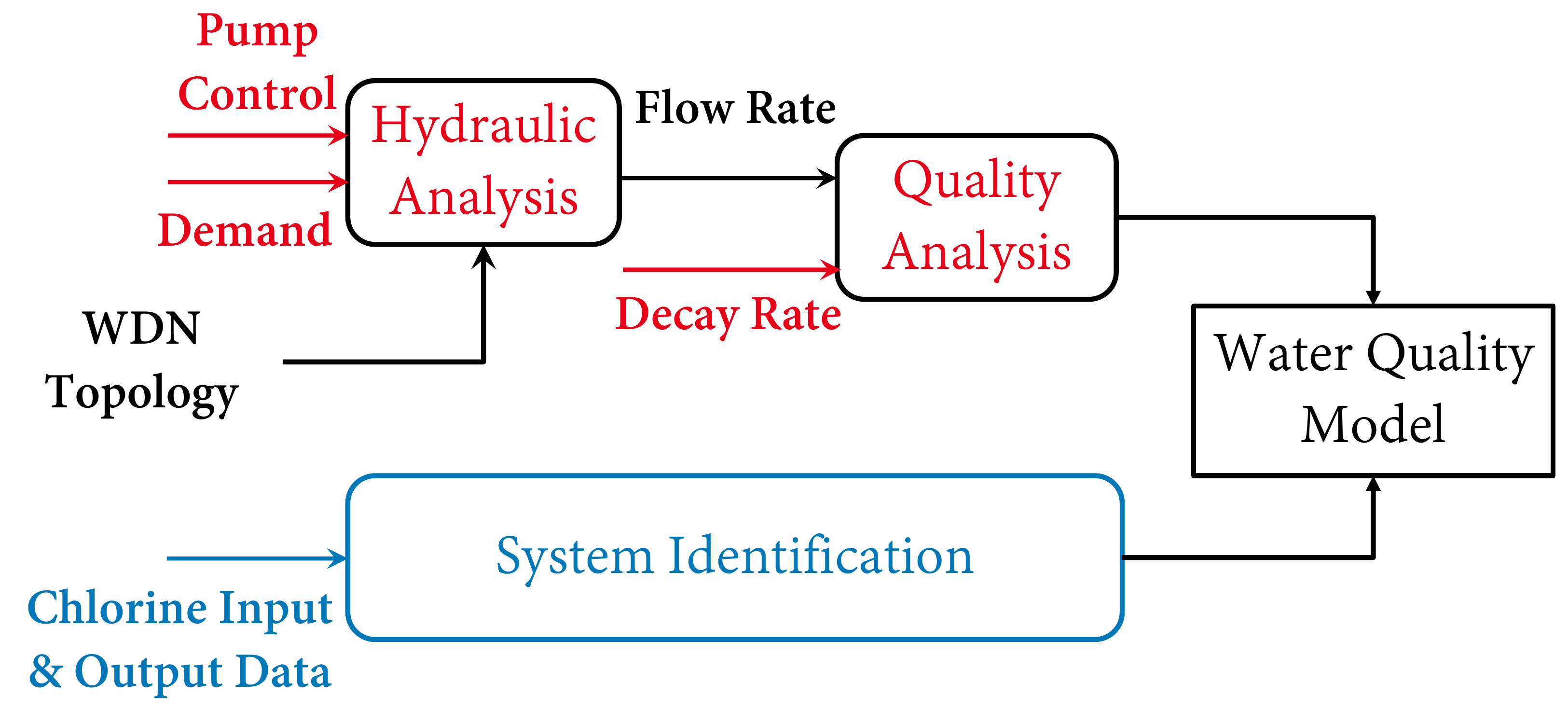}
	\caption{Two approaches to obtain a water quality model in a WDN---analysis-based and data-driven based methods.}
	\label{fig:input-output-dynamics}
\end{figure}

Many studies have investigated analysis-based water quality models~\cite{hua1999modelling,clark1994measuring,grayman1988modeling}, input-output models~\cite{zierolf1998development,shang2000input,polycarpou2002feedback,wang2005adaptive,duzinkiewicz2005hierarchical}, and control-oriented state-space model~\cite{wang2020effective}.  However, the analysis-based methods either rely on cumbersome, extensive water/chemistry based modeling or are based on obtaining hydraulic and many other parameters---or require both. For example, studies~\cite{zierolf1998development,shang2000input,wang2020effective} require a two-stage analysis to obtain the final water quality model, that is, water hydraulic analysis and water quality analysis (see Fig.~\ref{fig:input-output-dynamics}). This is because the water quality analysis depends on the hydraulic analysis results, such as the flow rates in all links. Furthermore,  all aforementioned studies have made more or less assumptions that simplify the true model such as single-species, first-order linear chlorine dynamics, a well-known WDN topology, and perfectly accurate parameters, while not considering uncertainty or leakage. Specifically, inputs of the analysis-based method, marked as red in Fig.~\ref{fig:input-output-dynamics}, contain uncertainties and are assumed as known. The limitations existing in analysis-based methods cannot be ignored, but fortunately, are avoided by data-driven based methods; see Table~\ref{tab:Comparison} for the comparisons.

A branch of dynamic system sciences, namely data-driven \textit{System Identification} (SysID) of dynamic system models, seeks to obtain dynamic models via data-driven approaches, without relying on advanced modeling or knowledge of network parameters or topology. The paper's objective is to thoroughly investigate, analytically and experimentally,  the performance of mainstream identification methods in computing dynamic state-space water quality models. To the best of our knowledge, this is the first study that seeks to analyze water quality dynamic models without relying on any network parameters or hydraulic simulation data and only utilizing input and output experimental data (e.g., the control data sequences applied to the system and measurements collected from installed sensors) with the help of using mainstream SysID algorithms.

\begingroup
\begin{table*}
	\caption{Comparison between analysis-based and data-driven based methods.}
	\centering
	\label{tab:Comparison}
	\renewcommand{\arraystretch}{1.4}
		\begin{tabular}{c|c|c|c|c|c|c}
			\hline
			\textit{Method}& \textit{Topology} & \textit{Parameter} & \textit{\makecell{Uncertainty \\ \& leakage}} & \textit{Explainable} & \textit{Hardware} & \textit{Realtime} \\ \hline
			\textit{\makecell{Analysis-based\\modeling}} & \makecell{Full\\topology} & \makecell{Accurate para-\\meters required} & \makecell{Sensitive \& \\hard to model} & Excellent & No needs & Yes \\ \hline
			\textit{\makecell{Data-driven \\based modeling}} & \makecell{No\\requirements} & \makecell{No\\requirements} & Robust & Poor & \makecell{Controllers\\ \& sensors} & Nearly\\ \hline \hline
		\end{tabular}
		\vspace{-0.4cm}
\end{table*}
\endgroup

\subsection{Literature review of system identification}~\label{sec:review}
The motivation of SysID is to construct system dynamics/models of a dynamical system from input-output experimental data. To this end, controllers and sensors are required to collect the necessary experimental data and corresponding well-designed algorithms are expected to estimate system dynamics. The models identified are found useful for control or estimation purposes; hence, SysID methods have been widely applied in industrial process~\cite{favoreel2000subspace}, aerospace~\cite{Juang1985,Horta1991}, and other fields such as chemistry~\cite{schaper1990identification}, as well as  computational fluid dynamics~\cite{schmid_2008,Brunton2016,vijayshankar2020dynamic}, and power grids~\cite{kamwa2000state}. 

Various SysID algorithms have been developed in the literature on dynamic systems. The majority of these algorithms can be divided into two categories: time-domain methods and frequency-domain methods. Time-domain methods are usually useful for complex, large-scale systems with many inputs and output measurements. Frequency-domain methods are popular in engineering practice for smaller systems~\cite{Shi2007}. The paper's focus is on time-domain methods as the water quality model in WDNs involves tens of thousands of states and tens of inputs and outputs~\cite{wang2020effective}. In what follows, we summarize the state-of-the-art of time-domain SysID algorithms. 

Time-domain methods can be categorized into various methods. Subspace identification methods (SIMs)~\cite{van1994n4sid,verhaegen1994identification,larimore1990canonical} estimate linear state-space models directly from time-discrete input-output experimental data and only need one parameter to be determined by users---the system order. The three famous SIMs from~\cite{van1994n4sid,verhaegen1994identification,larimore1990canonical} are known as canonical variate analysis (CVA), multivariable output-error state space (MOESP), and numerical subspace state space system identification (N4SID). Overschee and Moor~\cite{VanOverschee1995} propose a unifying theorem for these three SIMs. They point out that the three algorithms use the same subspace, but the weighting matrix, used to calculate a basis for the column space of the absorbability matrix, is slightly different.

In short, SIMs only assume a true underlying finite order linear time-invariant (LTI) system exists, and no a priori information about the system is needed. Moreover, SIMs are computationally attractive since only standard matrix operations such as singular value decomposition (SVD) are utilized, and iterative optimization is not involved. Hence, SIMs have been successfully applied to numerous multi-input multi-output (MIMO) applications~\cite{favoreel2000subspace,schaper1990identification}.

SysID algorithms can also be viewed as identifying a subspace spanned by an extended observability matrix, followed by the reconstruction of the corresponding state-space matrices~\cite{VanOverschee1995}. Therefore, methods implementing these two goals can be used to identify system dynamics. A method called eigensystem realization algorithm (ERA)~\cite{Juang1985,juang1986effects} is developed for modal parameter identification and model reduction of dynamic systems from impulse response. However, most realistic systems cannot meet the requirement of using impulse signals as inputs to obtain impulse response histories (i.e., Markov parameters). To solve such an issue, Juang \textit{et al.}~\cite{Juang1993} develop an observer/Kalman filter identification (OKID) that adopts commonly-used system inputs (e.g., step signal) to estimate the Markov parameters efficiently. Then, Markov parameters are passed to the ERA or some improved variants of it such as the eigensystem realization algorithm using data correlations (ERA/DC)~\cite{juang1987eigensystem} to complete the final SysID procedure~\cite{Vicario2014}. The approach combining OKID and ERA is often referred to as OKID/ERA, which is one of the most successful SysID algorithms. 

Recently, new system identification algorithms have been proposed such as dynamic mode decomposition (DMD) and sparse identification of nonlinear dynamics (SINDy). DMD~\cite{schmid_2008,Rowley2009} is a linear dynamic regression technique to map high-dimensional fluid data sequence into a dynamical system of significantly fewer degrees of freedom in computational fluid dynamics. 
Proctor \textit{et al.}~\cite{Proctor2016} further extend DMD to include inputs and control, and establish deeper connections to SIMs.  Bai \textit{et al.}~\cite{Bai2020} combine the compressed sensing DMD~\cite{brunton2015compressed} and DMD with input and control approaches~\cite{Proctor2016}, and propose a mathematical framework for compressive SysID. SINDy~\cite{Brunton2016}, another novel SysID approach, assumes the governing equations of nonlinear dynamics are sparse in the space of possible functions, and discovers governing equations from data by sparse identification. 

With so many years of development, SysID is a classical but still challenging research problem even for linear systems. For example, Zheng and Li~\cite{Zheng2020} focus on solving non-asymptotic identification of linear dynamical systems. Tsiamis and Pappas~\cite{Tsiamis2021} investigate when SysID is statistically challenging or less so. Using tools from minimax theory, their study shows that classes of linear systems (e.g., under-actuated or under-excited systems with weak coupling among the states) can be hard to identify. There is extensive and rich literature on theoretical and algorithmic developments of SysID, see~\cite{Ljung2010,Ljung2015,Qin2006,Favier2010,Zheng2020} for further details.

While the theoretical developments of SysID have been rich in the past decade, their application and investigation (analytically and experimentally) to challenging water network problems are rare. A related work from Polycarpou \textit{et al.}~\cite{polycarpou2002feedback} adopts input-output data to identify the unknown coefficients of the input-output model for water quality using adaptive parameter estimation, and the input-output model structure is identified with the help of hydraulic and water quality analysis from Zierolf \textit{et al.}~\cite{zierolf1998development}. However, the input-output model proposed in these two works differs from the state-space form proposed in this paper. To the best of our knowledge, this paper presents the first attempt to demystify SysID methods for estimating water quality dynamic models in the form of state-space representation using classical SysID methods.



\subsection{Paper contribution}~\label{sec:contribution}
The major objective of the paper is to thoroughly explore the applicability of SysID algorithms in water quality models, and investigate the corresponding performance and possible factors affecting the performance of SysID methods in estimating water quality dynamics. The below list outlines in detail the contributions of the paper. 
\begin{enumerate}
	\item We present the first attempt to identify water quality dynamics in state-space representation form in WDNs using only input-output data collected from experiments and classical SysID methods such as SIMs and ERA-based methods without knowing \textit{a priori} information about the system such as any parameters of WDNs.
	\item To adopt the suitable SysID methods and apply these algorithms correctly, exploring the properties or characteristics of water quality dynamics is important and necessary. Hence, we state and organize these significant and unique properties in WDNs such as \textit{linear-invariant},  \textit{high-delay} and \textit{overall uncontrollability and unobservability} from the control theory perspective. Moreover, the challenges caused by these properties when identifying water quality models (i.e., mismatches between SysID algorithms and the properties) are discussed, and possible solutions are proposed.
	\item Case studies show the potential of applying classical system identification algorithms in water quality models/dynamics, the performance of these algorithms in terms of accuracy and computational time are tested, and the possible factors affecting algorithm performance are explored. The significance of this work is that it allows water system operators to \textit{(i)} simulate water quality dynamics in the entire network without knowing network parameters, topology, or hydraulic and reaction parameters or even leaks, and \textit{(ii)} potentially apply modern control/estimation algorithms to manage water quality.
	
\end{enumerate}

The rest of the paper is organized as follows. Section~\ref{sec:WDNModel} briefly describes the basics of water quality modeling and its state-space representation. Section~\ref{sec:methods} introduces SysID algorithms potentially applicable/useful to water quality models. Properties of water quality models are proposed for the first time in Section~\ref{sec:discussion}, followed by the challenges caused by these properties when identifying water quality models and the discussion of possible solutions. Section~\ref{sec:casestudy} presents case studies showing the applicability of various SysID algorithms, their performance, and possible factors impacting the final results. Conclusions and future research directions are given in Section~\ref{sec:limitations}.

\noindent \textit{Paper's Notation.} \hspace{0.2cm} Boldface characters represent matrices and column vectors: $a$ is a scalar, $\m a$ is a vector, and $\m A$ is a matrix. Matrix $\m I$ denotes an identity square matrix, whereas $\m 0_{m \times n}$ denotes a zero matrix  with size $m$-by-$n$.
The notation $\mathbb{R}$ denotes the set of real and positive real numbers, and the notations $\mathbb{R}^n$ and $\mathbb{R}^{m\times n}$ denote a column vector with $n$ elements and an $m$-by-$n$ matrix in $\mathbb{R}$. For any vector $\m x \in \mathbb{R}^{n}$,  $\m x^{\top}$ is its  transpose. For any two matrices $\m A$ and $\m B$ with same number of columns, the notation $\{\m A, \m B\}$ denotes $[\m A^\top \  \m B^\top]^\top$. The induced norm of $\m A$ is $\norm {\m A}_2$, and the  Frobenius norm of $\m A$ is $\norm {\m A}_F$. The Moore-Penrose pseudoinverse of $\m A$ is denoted by $\m A^{\dagger}$.


\section{Water quality modeling}~\label{sec:WDNModel}
In this section, we succinctly present the fundamentals of water quality modeling considering single-species interactions and the corresponding state-space representation.  
\subsection{Water quality dynamics}\label{sec:dynamics}
We model WDN by a directed graph $\mathcal{G} = (\mathcal{N},\mathcal{L})$.  The set $\mathcal{N}$ defines the nodes and is partitioned as $\mathcal{N} = \mathcal{J} \cup \mathcal{T} \cup \mathcal{R}$ where $\mathcal{J}$, $\mathcal{T}$, and $\mathcal{R}$ are collection of junctions, tanks, and reservoirs. Let $\mathcal{L} \subseteq \mathcal{N} \times \mathcal{N}$ be the set of links, and define the partition $\mathcal{L} = \mathcal{P} \cup \mathcal{M} \cup \mathcal{V}$, where $\mathcal{P}$, $\mathcal{M}$, and $\mathcal{V}$ represent the collection of pipes, pumps, and valves. 

\begin{figure}[t]
	\centering
	\includegraphics[width=0.4\linewidth]{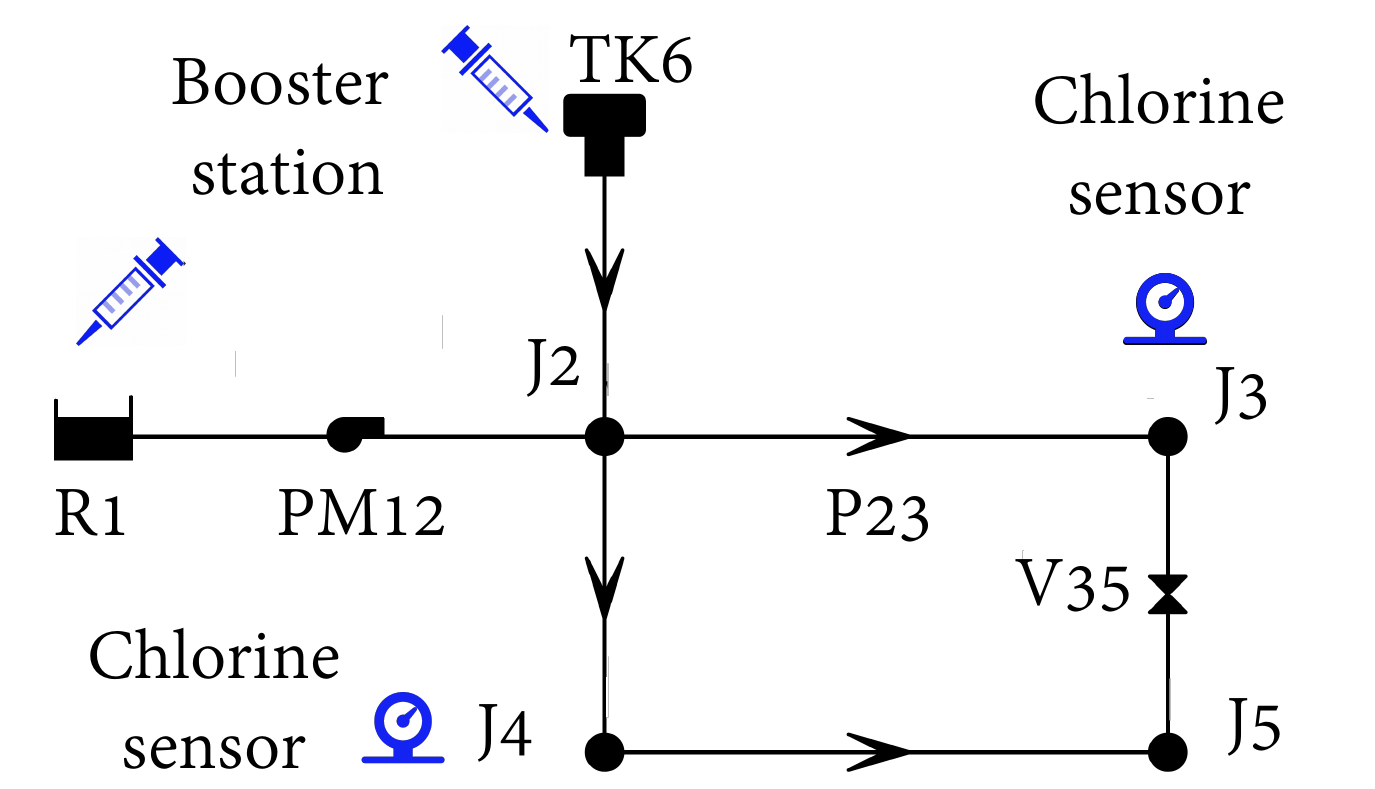}
	\caption{Exemplar topology of a WDN (multi-input and multi-output system with two boosters (inputs) and two sensors (outputs) to inject and to detect chlorine).}
	\label{fig:network}
\end{figure}


An example of a WDN graph is shown in Fig.~\ref{fig:network}, and a disinfectant (e.g., chlorine) is injected into the network with a proper mass and concentration by booster stations installed at the reservoir source R1 and the node TK6 to disinfect the WDN. Each junction in a WDN is a type of node that connects two links. Junctions may consume water, i.e., they have water demands which are assumed as known and unchanged during a hydraulic time-step (an hour or two) in water research. The concentration sensors are installed at specific nodes for measurement purposes (see the sensors installed at J3 and J4 in Fig.~\ref{fig:network}), in this way the water system operators can obtain the concentrations around the network. Ultimately, a water quality model representing the movement of all chemical and/or microbial species within a WDN as they traverse various components of the network can be represented by state-space form equations~\cite{wang2020effective}.

\subsection{State-space representation of water quality dynamics}


A water quality state-vector $\m x$ defining the concentrations of the disinfectant (chlorine) in the network at time $t$ is denoted as $\m x(t) \triangleq  \{\m c^\mathrm{N}(t),\m c^\mathrm{L}(t) \} \in \mbb{R}^{n_x}$ and $n_x = n_\mathrm{N} + n_\mathrm{L}$ where vectors $\m c^\mathrm{N}$ and $\m c^\mathrm{L}$ stand for the concentration at nodes and in links, the number of elements in vectors are $n_\mathrm{N}$ and $n_\mathrm{L}$. Specifically, pipes are split into segments, and each segment has a  variable in $\m c^\mathrm{L}$ standing for its concentration. After applying the principle of conversion of mass and assuming a constant chlorine decay rate,  we can rewrite all component models (i.e., water quality model) using a discrete-time state-space model of the form~\cite{wang2020effective}
\begin{equation}~\label{equ:fullmode-LTV}
	\begin{aligned}
		\m x(k+1) &= \m A(k)  \m x(k) + \m B(k)  \m u(k) \\
		\m y(k) &= \m C(k)  \m x(k) + \m D(k)  \m u(k),
	\end{aligned}
\end{equation}
where $\m x(k) \in \mathbb{R}^{n_x}$ is the state vector, the vectors $\m u(k) \in \mathbb{R}^{n_u}$ and $\m y(k) \in \mathbb{R}^{n_y}$ are the system input and output vectors (i.e., $n_u$ and $n_y$ are the number of booster stations and  sensors installed), and $\m A(k) $, $\m B(k) $, $\m C(k) $, and $\m D(k) $ are system matrices and time-varying. 

Note that system matrices are usually updated from half hour to several hours according to the changing rate of demands. That is, during a specific period which is referred to as \textit{hydraulic time-step}, the system matrices remain the same, and the discrete-time linear time-varying (DT-LTV) system~\eqref{equ:fullmode-LTV} degenerates into a discrete-time linear time-invariant (DT-LTI) system, that is
\begin{equation}~\label{equ:fullmode}
	\begin{aligned}
		\m x(k+1) &= \m A\m x(k) + \m B  \m u(k)\\
		\m y(k) &= \m C\m x(k) + \m D  \m u(k) .
	\end{aligned}
\end{equation}
Let us denote a generic $n_x$-th order DT-LTI system~\eqref{equ:fullmode} as the tuple $(\m A, \m B, \m C,\m D)$ for brevity.
In this way, the DT-LTV system~\eqref{equ:fullmode-LTV} comprises different tuples (DT-LTI systems). 
It is worthwhile to note that \textit{(i)}  water quality modeling is a DT-LTV system, \textit{(ii)} the DT-LTV system can be decomposed into several DT-LTI systems, and \textit{(iii)} we can apply SysID algorithms to each single DT-LTI system to obtain the identified model of the whole DT-LTV system.


\section{System identification methods}\label{sec:methods}
From the literature review, various SysID algorithms including SIMs, ERA-based, DMD-based, and SINDy methods can theoretically be applied to \textit{water quality model identification} (WQMI). However, we notice that not every method is suitable for our application. For example, DMD-based methods and SINDy are not applicable due to DMD assumes installing sensors at all nodes while SINDy depends on the choice of measurement variables, data quality, and the sparsifying function basis; this is thoroughly discussed in  Section~\ref{sec:Challengesolutions}.  Hence, only SysID algorithms that are potentially applicable to WQMI such as SIMs and ERA-based methods are introduced next. 

\subsection{Fundamentals of system identification}\label{sec:Fundamentals}
\subsubsection{Subspace identification method}  
Instead of introducing each algorithm of the SIMs, that are N4SID~\cite{van1994n4sid}, MOESP~\cite{verhaegen1994identification}, and CVA~\cite{larimore1990canonical}, we summarize the unifying theorem~\cite{VanOverschee1995} that unifies the three SIMs. Note that the identification problem in~\cite{VanOverschee1995} is considered in a combined deterministic-stochastic scenario. 
However, we only present the core ideas of the unifying SIM in a deterministic scenario for simplicity, since the key procedures considering state and measurement noises are similar for the stochastic scenario.

The unifying SIM organizes the state, input, and output data in a special way. Recall that $\m x(k)$ in~\eqref{equ:fullmode} stands for the system state, and it is collected into state sequence $\m X$, that is
\color{black}
\begin{equation}
\label{equ:x}
\m X =   \begin{bmatrix}
\m x(0)    &  \m x(1)   & \cdots   & \m x(k) & \m x(k+1) &  \cdots & \m x(k+m - 1)  &  \cdots
\end{bmatrix}.
\end{equation}

A sub-sequence $\m X_k \in \mathbb{R}^{n_x\times m}$ starting from index $k$ and includes $m$-step data of $\m X$ in \eqref{equ:x} is  
\begin{equation}\notag%
\m X_k =   \begin{bmatrix}
 \m x(k) & \m x(k+1) &  \cdots & \m x(k+m - 1)
 \end{bmatrix}.
\end{equation}

For example, when $k$ is set as $0$ in $\m X_k$, then $\m X_0$ is obtained.
 
Similarly, the $y(k)$ in ~\eqref{equ:fullmode} stands for the system output, and can be collected into $\m Y$, that is

\begin{equation}\notag%
\m Y =   \begin{bmatrix}
\m y(0)    &  \m y(1)   & \cdots   & \m y(k) & \m y(k+1) &  \cdots & \m y(k+m - 1)  &  \cdots
\end{bmatrix}.
\end{equation}

The output sub-sequence $\m Y_k \in \mathbb{R}^{n_y\times m}$ are defined similarly as $\m X_k$ as follows.

\begin{equation*}
\m Y_k =   \begin{bmatrix}
\m y(k) & \m y(k+1) &  \cdots & \m y(k+m - 1)
\end{bmatrix}
\end{equation*}

Furthermore, the input data sequence with $2k+m$ steps, that is $\m U = \begin{bmatrix}
	\m u(0) & \m u(1) & \cdots & \m u(2k+m-1) 
\end{bmatrix}$, is organized in the following form
\begingroup
\pgfkeys{tikz/mymatrixenv/.style={decoration={brace},every left delimiter/.style={xshift=8pt},every right delimiter/.style={xshift=-8pt}}}
\pgfkeys{tikz/mymatrix/.style={matrix of math nodes,nodes in empty cells,left delimiter={[},right delimiter={]},inner sep=1pt,outer sep=2pt,column sep=2pt,row sep=8pt,nodes={minimum width=5pt,minimum height=5pt,anchor=center,inner sep=2pt,outer sep=5pt}}}
\pgfkeys{tikz/mymatrixbrace/.style={decorate,thick}}
\pgfkeys{tikz/mymatrixbracenew/.style={decorate,thick,decoration={amplitude=8 pt}}}
\pgfkeys{tikz/mymatrixbracenewp/.style={decorate,thick,decoration={amplitude=4 pt}}}
\pgfkeys{tikz/mymatrixbracenewf/.style={decorate,thick,decoration={mirror,amplitude=4 pt}}}
\newcommand*\mymatrixbraceright[4][m]{
	\draw[mymatrixbrace] (#1.west|-#1-#3-1.south west) -- node[left=2pt] {#4} (#1.west|-#1-#2-1.north west);
}
\newcommand*\mymatrixbraceleft[4][m]{
	\draw[mymatrixbrace] (#1.east|-#1-#2-1.north east) -- node[right=2pt] {#4} (#1.east|-#1-#2-1.south east);
}
\newcommand*\mymatrixbracetop[4][m]{
	\draw[mymatrixbrace] (#1.north-|#1-1-#2.north west) -- node[above=2pt] {#4} (#1.north-|#1-1-#3.north east);
}
\newcommand*\mymatrixbracetopnew[4][m]{
	\draw[mymatrixbracenew] (#1.north-|#1-1-#2.north west) -- node[above=4pt] {#4} (#1.north-|#1-1-#3.north east);
}
\newcommand*\mymatrixbracetopnewp[4][m]{
	\draw[mymatrixbracenewp] (#1.north-|#1-1-#2.north west) -- node[above=3pt] {#4} (#1.north-|#1-1-#3.north east);
}
\newcommand*\mymatrixbracebottomf[4][m]{
	\draw[mymatrixbracenewf] (#1.south-|#1-1-#2.north west) -- node[below=2pt] {#4} (#1.south-|#1-1-#3.north east);
}
\newcommand*\mymatrixbracebottom[4][m]{
	\draw[mymatrixbrace] (#1.south-|#1-1-#2.north east) -- node[below=2pt] {#4} (#1.south-|#1-1-#3.north west);
}

\tikzset{
	style white/.style={
		set border color=black!90,fill opacity=0,
	},
	kwad/.style={
		above left offset={-0.52,0.52},
		below right offset={0.3,-0.3},
		#1
	},
	pion/.style={
		above left offset={-0.07,0.2},
		below right offset={0.07,-0.32},
		#1
	},
	fut/.style={
		above left offset={-0.52,0.4},
		below right offset={0.07,-0.52},
		#1
	},
	poz/.style={
		above left offset={-0.06,0.34},
		below right offset={0.1,-0.2},
		#1
	},
	paz/.style={
		above left offset={-0.1,0.25},
		below right offset={0.1,-0.3},
		#1
	},set fill color/.code={\pgfkeysalso{fill=#1}},
	set border color/.style={draw=#1}
}

\[
\m {U} = 
\begin{tikzpicture}[baseline={-0.5ex},mymatrixenv]
\matrix [mymatrix,inner sep=7pt,column sep=6pt, row sep=6pt] (m)  
{
 \m u(0)  &  \m u(1) &   \cdots   &   {\m u(k)}   & \cdots &  {\m u(m-1)}  &  \cdots &  \m u(k+m-1)  \\
	\m u(1)  & \m u(2) & \cdots & \m u(k+1) & \cdots & \m u(m) & \cdots & \m u(k+m) \\
	{ \vdots}  &  &  \ddots  &   & \vdots &    &  \ddots & \vdots  \\
	 {\m u(k-1)}    & {\m u(k)} &  \cdots  &  {\m u(2k-1)} & \cdots    & \m u(k+m-2)  & \cdots &   \m u(2k+m-1)   \\
};
\mymatrixbraceright{1}{1}{$\m{U}_0$}
\mymatrixbraceright{4}{4}{$\m{U}_{k-1}$}
\mymatrixbracetopnewp{1}{6}{$\m {U}_p$}
\mymatrixbracebottomf{4}{8}{$\m{U}_f$}
\mymatrixbraceleft{1}{1}{$\m{U}_{k}$}
\mymatrixbraceleft{4}{4}{$\m{U}_{2k-1}$}
\end{tikzpicture}
\]
\endgroup
\normalcolor
where input sub-sequences $\m U_0$ and $\m U_k$ are defined akin to $\m X_k$, whereas Hankel blocks $\m U_p$ and $\m U_f$ in $ \mathbb{R}^{(k n_x )\times (m n_u)}$ are considered as the past and future input data, that are $\left\lbrace \m U(0),  \cdots, \m U(k-1)\right\rbrace $ and $\left\lbrace \m U(k), \cdots, \m U(2k-1)\right\rbrace$.

System identification from unifying SIM~\cite{VanOverschee1995} is achieved by solving a least-square problem, that is
\begin{equation}~\label{equ:leastsquare}
	\underbrace{\begin{bmatrix}
			\m X_{k+1} \\
			\m Y_k
	\end{bmatrix}}_{\text{known}}= 	\begin{bmatrix}
		\m A & \m B \\
		\m C & \m D
	\end{bmatrix} \underbrace{\begin{bmatrix}
			\m X_k \\
			\m U_k
	\end{bmatrix}}_{\text{known}},
\end{equation}
where input $\m U_k$ and output $\m Y_k$ data sequences are known, and if  $\m X_k$ and $\m X_{k+1}$ which are also referred to as \textit{Kalman state sequence} can be obtained, then system matrices  $(\m A, \m B, \m C, \m D)$ can be fully recovered from~\eqref{equ:leastsquare} by the least-square method. Next, we briefly present how unifying SIM estimates Kalman state sequence $\m X_k$ or $\m X_{k+1}$.

With defining data sequences such as $\m X_k$, $\m Y_k$, $\m U_p$, and $\m U_f$, the system~\eqref{equ:fullmode} can be rewritten as
\begin{equation}~\label{equ:agg}
	\begin{aligned}
		\m X_k &= \m A^k \m X_0 + \m \Delta_k  \m U_p \\
		\m Y_0 &= \m \Gamma_k  \m X_0 + \m H_k  \m U_p \\
		\m Y_k &= \m \Gamma_k  \m X_k + \m H_k  \m U_f,
	\end{aligned}
\end{equation}
where 	extended observability  matrix $\m \Gamma_k$ and Toeplitz matrix $\m H_k$ are defined by\\
\begingroup
\small
\setlength\arraycolsep{0.5pt}
\setlength{\tabcolsep}{1.2pt}
\begin{subequations}
	\noindent    \begin{tabularx}{\linewidth}{m{0.39\linewidth}m{0.59\linewidth}}
		\begin{equation}
			\label{equ:gamma}
			\m \Gamma_k = \begin{bmatrix}
				\m C \\
				\m C \m A\\
				\vdots\\
				\m C \m A^{k-1}
			\end{bmatrix},
		\end{equation}
		&
		\begin{equation}
			\label{equ:hs}
			\m H_k =   \begin{bmatrix}
				\m D &   &   &  \\
				\m C \m B & \m D &   &  \\
				\vdots &   & \ddots &  \\
				\m C \m A^{k-2} \m B &    & \cdots & \m D
			\end{bmatrix} .
		\end{equation}
	\end{tabularx}
\end{subequations}
\endgroup

If one observes the three subequations in~\eqref{equ:agg} carefully, the connections among input and ouput data sequences $\m Y_k$, $\m Y_0$, $\m U_p$, and $\m U_f$ are achieved by state sequences $\m X_k$, $\m X_0$, and other necessary matrices. The unifying SIM theorem points out that  matrix $\mc{\m O}_k = \m \Gamma_k  \m X_k$ is equal to the oblique projection of $\m Y_k$ along $\m U_f$ direction on the past data $\begin{bmatrix}
	\m U_p \\\m Y_0
\end{bmatrix}$, and each column of it is the autonomous response to initial states; see Appendix in~\cite{VanOverschee1995} for more details. In short, $\mc{\m O}_k = \m \Gamma_k  \m X_k$ representing autonomous response can be easily found by calculating an oblique projection. Moreover, Kalman state sequence $\m X_k$ which is our goal can be extracted directly from SVD for $\mc{\m O}_k$, that is 
\begingroup
\setlength\arraycolsep{1pt}
\begin{equation} ~\label{equ:svdofO}
	\mc{\m O}_k =  \m U \m \Sigma \m V^{\top} = \underbrace{\m U \m \Sigma^{\frac{1}{2}}}_{\textstyle \m \Gamma_k}  \underbrace{\m \Sigma^{\frac{1}{2}}\m V^{\top}}_{\textstyle \m X_k}.
\end{equation}
\endgroup

We can adopt similar procedures to estimate Kalman state sequence $\m X_{k+1}$, and after substituting them into~\eqref{equ:leastsquare}, system matrices can be identified. In addition to~\eqref{equ:leastsquare}, another popular strategy to recover system matrices is introduced in~\cite{Verhaegen1994} by taking advantage of the special structure of $\m \Gamma_k$ and the special Toeplitz structure of $\m H_k$. 

The aforementioned procedures are only the core idea of unifying SIM, other details are simplified to help readers understand the big picture of SIMs. For example,~\eqref{equ:svdofO} is actually a special case of SVD of $\m W_1 \mc{\m O}_k \m W_2$ when setting weight matrices $\m W_1 = \m W_2 = \m I$. In this way, this unifying method generates the same results as the N4SID method published in~\cite{van1994n4sid}.  When selecting different matrices $\m W_1$ and $\m W_2$ combinations, the unifying SIM can be viewed as CVA and MOESP methods as well. Moreover, to reduce the system order, we can choose the first $n_r$ singular values when performing SVD for $\m W_1 \mc{\m O}_k \m W_2$, that is
\begingroup
\setlength\arraycolsep{1.2pt}
\setlength{\tabcolsep}{1pt}
\begin{equation}~\label{equ:svd-o}
	\hspace{-1em} \m W_1 \mc{\m O}_k \m W_2 =  \m U \m \Sigma \m V^{\top} = \underbrace{\begin{bmatrix}
			\m U_r & \m 0
		\end{bmatrix} \begin{bmatrix}
			\m \Sigma_r^{\frac{1}{2}} & \m 0\\
			\m 0 & \m 0
	\end{bmatrix}}_{\textstyle \m \Gamma_k}  \underbrace{\begin{bmatrix}
			\m \Sigma_r^{\frac{1}{2}} & \m 0\\
			\m 0 & \m 0
		\end{bmatrix}\begin{bmatrix}
			\m V_r^{\top} \\ \m 0 
	\end{bmatrix}}_{\textstyle \m X_k}.
\end{equation}
\endgroup
As for how to determine a proper value for system order $n_r$, the users could use trial-and-error method or a binary search based algorithm provided at the end of Section~\ref{sec:algorithm}; see Algorithm~\ref{alg:greedy} for detail.
The unifying SIM are summarized as Procedure~\ref{proc:unified}.
\begin{procedure}
	\small \DontPrintSemicolon
	\KwIn{Input block Hankel matrices $\m U_{p}$ and $\m U_{f}$, ouput data sequences $\m Y_{0}$ and $\m Y_{k}$, system order $n_r$}
	\KwOut{Identified system matrices  $(\m A, \m B, \m C, \m D)$}
	Solve an oblique projection from $\m Y_{k}$ to  $\begin{bmatrix}
		\m U_p \\\m Y_0
	\end{bmatrix}$ along $\m U_f$ direction and obtain autonomous response  $\mc{\m O}_k$\;
	Select proper weight matrices $\m W_1$ and $\m W_2$  in~\eqref{equ:svd-o}  to determine the use of CVA, MOESP, or N4SID\; 
	Choose system order $n_r$ and perform  SVD for $\m W_1 \mc{\m O}_k \m W_2$ as ~\eqref{equ:svd-o} to obtain Kalman state sequence $\m X_{k}$, $\m X_{k+1}$ in~\eqref{equ:leastsquare}, and extended observability matrix $\m \Gamma_k$ in~\eqref{equ:gamma} \; 
	Obtain $(\m A, \m B, \m C, \m D)$ by~\eqref{equ:leastsquare} or the strategy in~\cite{Verhaegen1994}\;
	\caption{Unifying subspace identification method() (SIM)}
	\label{proc:unified}
\end{procedure}

\subsubsection{ERA method} As mentioned in Section~\ref{sec:review}, ERA~\cite{Juang1985} is developed for a finite-dimensional, DT-LTI dynamical system such as~\eqref{equ:fullmode}. The core idea is to use impulse signals as system inputs to construct Markov parameters, that are $\m D$ and $\m C \m A^{k-1} \m B$. Specifically, suppose that the unit impulse signal is defined as
\begin{equation}\notag
	u_i(k) =  \begin{dcases}
		1, k = 0\\
		0, k \neq 0
	\end{dcases},
\end{equation}
where input $u_i(k)$ is the element of $\m u$ with index of $(i,k)$.

Then, the output sequence can be expressed by  \begin{align} \label{equ:impulseresponse}
	\m y(k) =  \begin{dcases}
		\m D, \hspace{3.2em}k = 0\\
		\m C \m A^{k-1} \m B, k \neq 0
	\end{dcases}.
\end{align}
Note that matrix $\m D$ in~\eqref{equ:fullmode} is already identified and can be obtained directly from $\m y(0)$. We present how to identify other matrices next. 

First, the ERA algorithm begins by forming the \textit{block Hankel matrix} $ \m H_m$:
\begingroup
\small
\setlength\arraycolsep{2pt}
\begin{equation}~\label{equ:hankel}
	\m H_m =   \begin{bmatrix}
		\m y(1) & \m y(2) & \cdots & \m y(m_c)\\
		\m y(2) & \m y(3) & \cdots & \m y(m_c+1)\\
		\vdots & \vdots & \ddots & \vdots\\
		\m y(m_o) & \m y(m_o+1) & \cdots & \m y(m_c + m_o)
	\end{bmatrix}  
	= \begin{bmatrix}
		\m C \m B & \m C \m A \m B & \cdots & \m C \m A^{m_c} \m B\\
		\m C \m A \m B & \m C \m A^{2} \m B & \cdots & \m C \m A^{m_c + 1} \m B\\
		\vdots & \vdots & \ddots & \vdots\\
		\m C \m A^{m_o}  \m B & \m C \m A^{m_c + 1}  \m B  &\cdots & \m C \m A^{m_c + m_o} \m B
	\end{bmatrix}, \notag
\end{equation}
\endgroup
where parameters $m_c$ and $m_o$ are the numbers of steps.

Second, ERA computes SVD of $\m H_m$ in~\eqref{equ:svd} to obtain $\m U_r $ and $\m V_r$.
\begingroup
\setlength\arraycolsep{1pt}
\begin{equation}~\label{equ:svd}
	\m H_m =  \m U \m \Sigma \m V^{\top} = \begin{bmatrix}
		\m U_r & \m 0
	\end{bmatrix} \begin{bmatrix}
		\m \Sigma_r & \m 0\\
		\m 0 & \m 0
	\end{bmatrix} \begin{bmatrix}
		\m V_r^{\top} \\ \m 0 
	\end{bmatrix},
\end{equation}
\endgroup
where singular value $\m \Sigma_r$ is in $n_r \times n_r$, and constant $n_r$ is chosen by users as the dimension of identified system or obtained by Algorithm~\ref{alg:greedy} at the end of Section \ref{sec:algorithm}.

Finally, the identified system matrices $(\m A, \m B, \m C, \m D)$ are
\begin{equation}\label{equ:ERARESULT}
	\begin{aligned}
		\m  A &=\m \Sigma_{r}^{-\frac{1}{2}} \m  U_{r}^{\top} \m  H_m^{\prime} \m V_{r} \m  \Sigma_{r}^{-\frac{1}{2}},\\
		\m  B &= \text{the first } n_u \text{ columns of } \m  \Sigma_{r}^{\frac{1}{2}} \m  V_{r}^{\top},\\
		\m  C &= \text{the first } n_y \text{ rows of }   \m  U_{r} \m \Sigma_{r}^{\frac{1}{2}},\\
		\m D &= \m y(0),
	\end{aligned}
\end{equation}
where 
\begingroup
\small
\setlength\arraycolsep{1pt}
\begin{equation}~\notag
	\m H_m^\prime =   \begin{bmatrix}
		\m C \m A \m B & \m C \m A^{2} \m B & \cdots & \m C \m A^{m_c + 1} \m B\\
		\m C \m A^2 \m B & \m C \m A^{3} \m B & \cdots & \m C \m A^{m_c + 2} \m B\\
		\vdots & \vdots & \ddots & \vdots\\
		\m C \m A^{m_o + 1}  \m B & \m C \m A^{m_c + 2}  \m B & \cdots & \m C \m A^{m_c + m_o  +2} \m B
	\end{bmatrix}.
\end{equation}
\endgroup

The general ERA procedure from~\cite{Juang1985} are summarized as Procedure~\ref{proc:ERA}.

\begin{procedure}
	\small \DontPrintSemicolon
	\KwIn{Impulse signal, \textcolor{black}{system order $n_r$}}
	\KwOut{Identified system matrices  $(\m A, \m B, \m C, \m D)$}
	Collect output data using \eqref{equ:input-ouput}.
	Obtain Markov parameters $\m D$ and $\m C \m A^{k-1} \m B$ using impulse signal\;
	Construct block Hankel matrix $\m H_m$ using \eqref{equ:hankel}\;
	\textcolor{black}{Perform  SVD of $\m H_m$ via  \eqref{equ:svd} with chosen system order $n_r$}\;
	Obtain $(\m A, \m B, \m C, \m D)$ by~\eqref{equ:ERARESULT}\;
	\caption{Eigensystem realization algorithm() (ERA)}
	\label{proc:ERA}
\end{procedure}
Note that ERA requires impulse signal as system inputs, and this may be not applicable to  WQMI since the amplitude of impulse response is not large enough to be detected by sensors. We investigate that in the ensuing section and the case study.


\subsubsection{OKID/ERA method} 
To avoid using the impulse signal and make ERA more general, OKID~\cite{Juang1993} is developed to efficiently estimate the Markov parameters (i.e., $\m D$ and $\m C \m A^{k-1} \m B$). With an assumption of zero-initial conditions, the system equation~\eqref{equ:fullmode} can be rewritten as
\begingroup
\small
\setlength\arraycolsep{1.4pt}
\renewcommand{\arraystretch}{1.4}
\begin{equation}\label{equ:input-ouput}
	\underbrace{\begin{bmatrix}
			\m y(0) \\
			\m y(1) \\
			\vdots\\
			\m y(m)
	\end{bmatrix} }_{\textstyle \m y_m}	
	= 	
	\underbrace{\begin{bmatrix}
			\m D & \m C \m B & \cdots & \m C \m A^{m} \m B
	\end{bmatrix}}_{\textstyle \m Y_m}	
	\underbrace{\begin{bmatrix}
			\m  u(0) & \m  u(1) & \cdots  & \m  u(m)\\
			& \m  u(0) & \cdots  & \m  u(m-1)\\
			&  & \ddots & \vdots  \\
			& &  & \m  u(0)
	\end{bmatrix}}_{\textstyle \m U_m },
\end{equation}
\endgroup
where parameter $m$ represent the number of steps. All Markov parameters are collected in $\m Y_m$ that connects $m-$step  input matrix $\m U_m$ and output sequence $\m y_m$.

Then, the Markov parameters located in $\m Y_m$ can be simply solved by $\m y_m \m U_m^{\dagger}$. As for large-scale systems or a large $m$, $\m Y_m$ can also be solved by SVD techniques.  After Markov parameters are obtained, the standard ERA procedure is called to complete the final system identification; the general OKID/ERA procedure is presented in Procedure~\ref{proc:ERAOKID}.

\begin{procedure}
	\small \DontPrintSemicolon
	\KwIn{System input sequence $\m u(0), \m u(1), \cdots, \m u(m)$; output sequence $\m y(0), \m y(1), \cdots, \m y(m)$,  \textcolor{black}{system order $n_r$}}
	\KwOut{Identified system matrices  $(\m A, \m B, \m C, \m D)$}
	Construct input-output relationship \eqref{equ:input-ouput} and
	solve for each Markov parameter in matrix $\m Y_m$ \;
	Call Steps 2-4 in Procedure~\ref{proc:ERA} to  obtain $(\m A, \m B, \m C, \m D)$\;
	\caption{OKID followed by eigensystem realization algorithm() (OKID-ERA)}
	\label{proc:ERAOKID}
\end{procedure}

\subsection{System order determination algorithm }\label{sec:algorithm}

As mentioned in the above procedures, system order $n_r$ is the only parameter to be determined by users. The users could finish this task according to their experiences and adopt the trial-and-error method. However, setting system order by intuition is cumbersome and is not a stable way when the singular values vary significantly for different matrices to be decomposed using the SVD method. Hence, we provide a stable indicator (i.e., energy level) for determining system order next.

Singular values define the``energy” of each state in a system $(\m A, \m B, \m C,\m D)$. Thus, keeping larger energy states of a system preserves most of its characteristics in terms of stability, frequency, and time responses. Note that each SysID algorithm mentioned in this paper includes SVD procedure and an available singular value matrix $\m \Sigma$; see~\eqref{equ:svd-o} or~\eqref{equ:svd}. To determine the system order automatically and efficiently, we use the concept of energy level---the ratio of the sum of the $n_r$ largest  singular values to the sum of all singular value---that is
\begin{equation}~\label{equ:el}
\m l = \dfrac{\sum_{i=1}^{n_r} \m \Sigma_{i,i}}{\sum_{i=1}^{n} \m \Sigma_{i,i}}.
\end{equation}
To preserve most of the characteristics in a system, the users simply set a minimal energy level goal (e.g., $l > 95\%$) which corresponds to a certain $n_r$. We provide a binary search algorithm to determine the system order $n_r$ when the energy level is set by users, and the detailed steps are given in Algorithm~\ref{alg:greedy}.
\begin{algorithm}
	\small	\DontPrintSemicolon
	\KwIn{Energy level goal $g$, singular value matrix $\m \Sigma$  in \eqref{equ:svd-o} or~\eqref{equ:svd}}
	\KwOut{System order $n_r$}
	{Let}  $n \leftarrow$ number of elements in singular value matrix $\m \Sigma$, $left \leftarrow 1$, $right
	 \leftarrow n$  \;
	 {Compute} the sum of singular values $S = \sum_{i=1}^{n} \m \Sigma_{i,i}$\;
	\While{$\lvert left - right \rvert > 1$}{
		$n_r \leftarrow \left\lfloor \frac{left + right}{2} \right\rfloor$
		\hspace{9em}   //  Find the middle number as current order\;	
		{Obtain} current energy level $l \leftarrow \sum_{i=1}^{n_r} \m \Sigma_{i,i}/S$ \hspace{1.27em} // Find the current energy level\;	
		\eIf{$l > g$}{
			$right \leftarrow n_r$\;
		}{
			$left \leftarrow n_r$\;
		}
	}
	\caption{Binary search based algorithm to determine system order $n_r$.}
	\label{alg:greedy}
\end{algorithm}

\section{Properties of water quality model, challenges of identification, and possible solutions}~\label{sec:discussion}
Unlike other applications, the water quality model has special characteristics/properties. 
Hence, we show these special properties, present challenges caused by them, and discuss corresponding solutions while performing WQMI. 

To illustrate these issues clearly, we use the network comprised of an output sensor and a booster station controller as an example; see Fig.~\ref{fig:systemidentification}. As the aforementioned topology in Section~\ref{sec:dynamics}, the WDN in Fig.~\ref{fig:systemidentification} is expressed by a directed graph $\mathcal{G} = (\mathcal{N},\mathcal{L})$. To simplify, only junctions are included in set $\mathcal{N}$, and the set $\mathcal{L}$ only collects pipes in this example. Furthermore, the amount of chlorine injected by the booster station controller in Fig.~\ref{fig:systemidentification} is viewed as the system input data; the chlorine concentration measured by the sensor in Fig.~\ref{fig:systemidentification} is the corresponding system output data.

\subsection{Properties of water quality model}
Let us present the process of collecting the input-output data for SysID algorithms at first, and then the special properties can be summarized from the process intuitively.

\begin{figure}[t]
	\centering
	\includegraphics[width=0.5\linewidth]{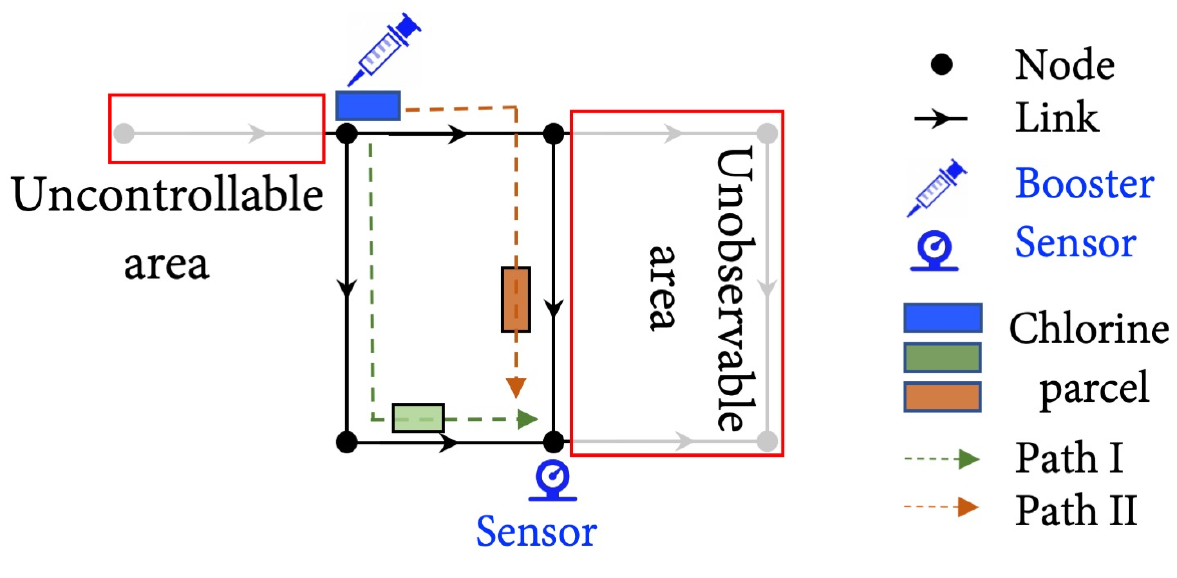}
	\caption{An example to illustrate the properties of water quality dynamics.}
	\label{fig:systemidentification}
\end{figure}

During a fixed hydraulic time-step, junction demands remain the same, the hydraulics in a WDN are unchanged, and the water quality model can be described as a DT-LTI system~\eqref{equ:fullmode}.  After a chlorine parcel with a certain mass and concentration is injected at the booster station (i.e., the blue parcel in Fig.~\ref{fig:systemidentification}),  it travels to the sensor with various velocities in different links (i.e., the Paths I and II in Fig.~\ref{fig:systemidentification}). During traveling, both mass and concentration of the chlorine parcel reduce gradually due to consumption by junctions and decay of chlorine. Furthermore, parcels with different masses and concentrations (see the light green and brown parcels in Fig.~\ref{fig:systemidentification}) arrive at the sensor at various times. Data collection starts in all booster stations and sensors from the moment when the booster station injects the chlorine parcel, and stops at the moment when the last chlorine parcel reaches the sensor.

From the above input-output data collection process, we note several properties of the water quality model below from the control theory perspective.

\begin{itemize}
	\item The water quality model is a special DT-LTV system with regular and consistent properties. That is, it degenerates into a simple DT-LTI system during each hydraulic time-step.  We refer this property as \textit{linear-invariant property}.
	
	\item In practice, the water quality model is stable. This \textit{stable property} is intuitive because the injected chlorine would gradually decay and disappear eventually in a WDN after enough time. A system operator continuously injects chlorine to ensure that the chlorine concentrations do not decay to zero, which would violate federal water quality mandates. 
	

	\item The overall water quality model, in general, is uncontrollable and unobservable from a control-theoretic perspective due to the limited number of booster stations (i.e., controllers) and measurement sensors that are allowed to be installed in many real-world WDNs. That is, there exist certain components that cannot be affected by booster stations, and certain components have no effect on the sensors; see the uncontrollable and unobservable areas marked by red, rectangular boxes in Fig.~\ref{fig:systemidentification}. This property is referred to as \textit{overall uncontrollability and unobservability}.

	\item Significant delays exist in the water quality model. The delay, caused by the slow traveling time from booster stations to sensors, is a natural attribute of the water supply system. This property is named as \textit{high-delay property}.
	
	\item Senors are assumed to be only installed at nodes in reality and installing sensors in links (e.g., in the center of a pipe) is not allowed. This  \textit{partial-state access property} indicates that $\m C$ is not allowed to be $\m I$. That is, full-state access is impossible in a water quality scenario.
	
\end{itemize}

\subsection{Challenges and solutions for water quality model identification}~\label{sec:Challengesolutions} From the above property analysis, we note that the \textit{linear-invariant} and \textit{stable}  properties are supportive and friendly for WQMI using SysID algorithms since most SysID algorithms are designed for stable DT-LTI systems.
As for WQMI for a DT-LTV system, that is identifying quality models when demand pattern changes in an extended period simulation (e.g., 24 hours), we note that demands changing leads to the changes of flow rates that would cause the switch of subsystems. However, the determination algorithm (i.e., Algorithm~\ref{alg:greedy}) would remain applicable. Hence, the value of $n_r$ may vary in different hydraulic time-steps, and system matrices $(\m A, \m B, \m C, \m D)$ with different system orders are obtained.

However, the \textit{overall uncontrollability and unobservability}, \textit{high-delay}, and \textit{partial-state access} properties would cause problems and still are challenges for current state-of-the-art SysID algorithms.

First, SIMs introduced in this paper assume a controllable and observable system~\cite{larimore1990canonical,verhaegen1994identification,van1994n4sid,VanOverschee1995} and  ERA-related methods~\cite{Juang1985,Juang1993} also make similar assumptions. Although standard DMD has no such requirements, it assumes full-state access and cannot deal with control inputs. Compressed sensing DMD~\cite{Brunton2013}  and random sampling strategy DMD~\cite{BenjaminErichson2019} require a special measurement matrix $\m C$. For example,  the $\m C$ must be incoherent to the sparse basis~\cite{Brunton2013}. It indicates DMD-related methods are not applicable. Hence, the \textit{overall uncontrollability and unobservability} property indicates that the SysID of water quality model for overall WDN is infeasible.

Nevertheless, there are always both controllable and observable areas in a WDN, that is, the part ``surrounded" by controllers and sensors (see the middle area in Fig.~\ref{fig:systemidentification}). The area is controllable because the chlorine concentrations of all pipes and junctions in that area can be affected by the booster station; all chlorine concentrations have effects on the sensors indicating the existence of observable property. From a control-theoretical perspective, the state-space representation of the middle part is the controllable and observable subsystem after performing Kalman decomposition for the overall/full state-space representation. Therefore, we are allowed to adopt the current popular SysID algorithms to estimate that part of the water quality model in a WDN. Moreover, the identified subsystem would have the same behavior as the part of the original full system surrounded by controllers and sensors.

Second, the \textit{high-delay property} also brings challenges to WQMI, which is reflected in the completeness of the data received by sensors. Suppose that the traveling time from the booster station controllers to sensors is larger than the hydraulic time-step. Sensors have not yet received enough/complete data for SysID, and the current DT-LTI system has been switched to the next one because of the updated demand. This would cause a direct failure for  SysID algorithms. Hence, the duration of the signal selected as input has to be as short as possible or finite length such that the data completeness can be guaranteed. 

There are two possible solutions for this high-delay issue. The simple one is to install the controllers and sensors close enough to each other so that the traveling time is always less than a hydraulic time-step. But this clearly would limit the size of the system we can identify. Moreover, this might be infeasible since the system operator might not have enough budget to install more sensors or controllers. The complicated one is to use repeaters to cascade the system when the controllers and sensors are far away from each other. The repeaters are actually sensors but viewed as system inputs. That is, the chlorine concentration data measured by repeater sensors is treated as the input data injected by booster stations. In this way, a large-scale system that cannot be identified due to a lack of data completeness is solved by cascading two identified subsystems.   Suppose we have two identified small-scale subsystems $(\m A_1, \m B_1,\m C_1,\m D_1)$ and $(\m A_2, \m B_2,\m C_2,\m D_2)$. After cascading, the final identified system is 
\begin{equation}\notag
\left( \begin{bmatrix}
\m A_1 & \\
\m B_2 \m C_1 & \m A_2
\end{bmatrix}, \begin{bmatrix}
\m B_1\\
\m B_2 \m D_1
\end{bmatrix},\begin{bmatrix}
\m D_2 \m C_1 & \m C_2
\end{bmatrix}, \m D_2 \m D_1\right).
\end{equation}

Third, the challenge from \textit{partial-state access property} is tricky. If WQMI for the overall system is expected, then there is no solution since installing sensors at all junctions and in all links is impossible to meet the controllable and observable requirements in most SysID algorithms. However, suppose the goal is only to identify the subsystem discussed in the aforementioned \textit{overall uncontrollability and unobservability property} instead of the whole system, then the $\m C$ for the overall system does not have to be the identity matrix, meaning that the subsystem can be identified using only a few sensors installed. 

\subsection{Further discussions of the applicability of proposed WQMI methods}
Although challenges and possible solutions for WQMI are presented in the previous section, several crucial points, such as the number of controllers and sensors and possible ways to apply the proposed method in practical large-scale networks, are not given and are discussed next.

In a practical large-scale WDN, our method is applicable for WQMI in several specific zones that do not overlap (i.e., zone-level identification) since it can be simplified into identifying several individual LTI subsystems. In such a case, only a limited number of controllers and sensors are needed. That is, at least one booster and one sensor are needed for each zone. However, if all components' quality models need to be identified (i.e., component-level identification), the controllers and sensors must be installed almost at all nodes.

Furthermore, we note that the furthest distance or the equivalent traveling time between controllers and sensors in each zone should be less than or equal to the hydraulic time-step in an extended period simulation if repeaters are not used. When the traveling time is equal to or slightly less than a full hydraulic time-step, it indicates a quality model is identified, and a new hydraulic time-step immediately arrives. It seems that the identified model has to be discarded and is useless. However, the identified model is still helpful for water quality management, and we present a potential scenario below.

The nodal demand changes each hydraulic time-step in a zone of WDNs. However, the demand pattern in the zone remains similar or stable each day of the week. It indicates that the water quality model is similar during a specific time at the zone. In such a way, we can adopt model predictive control (MPC) to manipulate the chlorine concentrations in that area/zone since we already have the identified model from the previous day, and MPC is well-known for its effectiveness even when the model it adopts is not accurate.

Moreover, the focus of this paper is to show the potential to use mainstream system identification algorithms for WQMI rather than to show how to use the identified model. As for how to combine WQMI with the MPC method and apply them to solve water management issues in practical large-scale networks, we will explore such topics in future work.

\section{Case Studies}~\label{sec:casestudy}
We present two examples (i.e., Three-node and Net1  networks~\cite{shang2002particle}) to illustrate the properties (e.g., stability) and the performance of different SysID methods in terms of accuracy and computational time. 
The testing environment is an Ubuntu 16.04 Precision 5810 Tower with Intel(R) Xeon(R) CPU E5-1620 v3 @ 3.50GHz and 8G. Specially, in this numerical study, we attempt to answer the following questions:
\begin{itemize}
	\item[-] Q1: Can the water system operator only utilize the collected input-output experimental data to identify the water quality model without knowing any other network parameters?
	\item[-] Q2: How do the aforementioned algorithms perform in terms of performance (e.g., accuracy and running time) of tested SysID methods for different networks?
	\item[-] Q3: How can SysID methods help water system operators to control the chlorine concentration in the interested area?

\end{itemize}

\begin{figure}[t]
	\centering
	\includegraphics[width=0.57\linewidth]{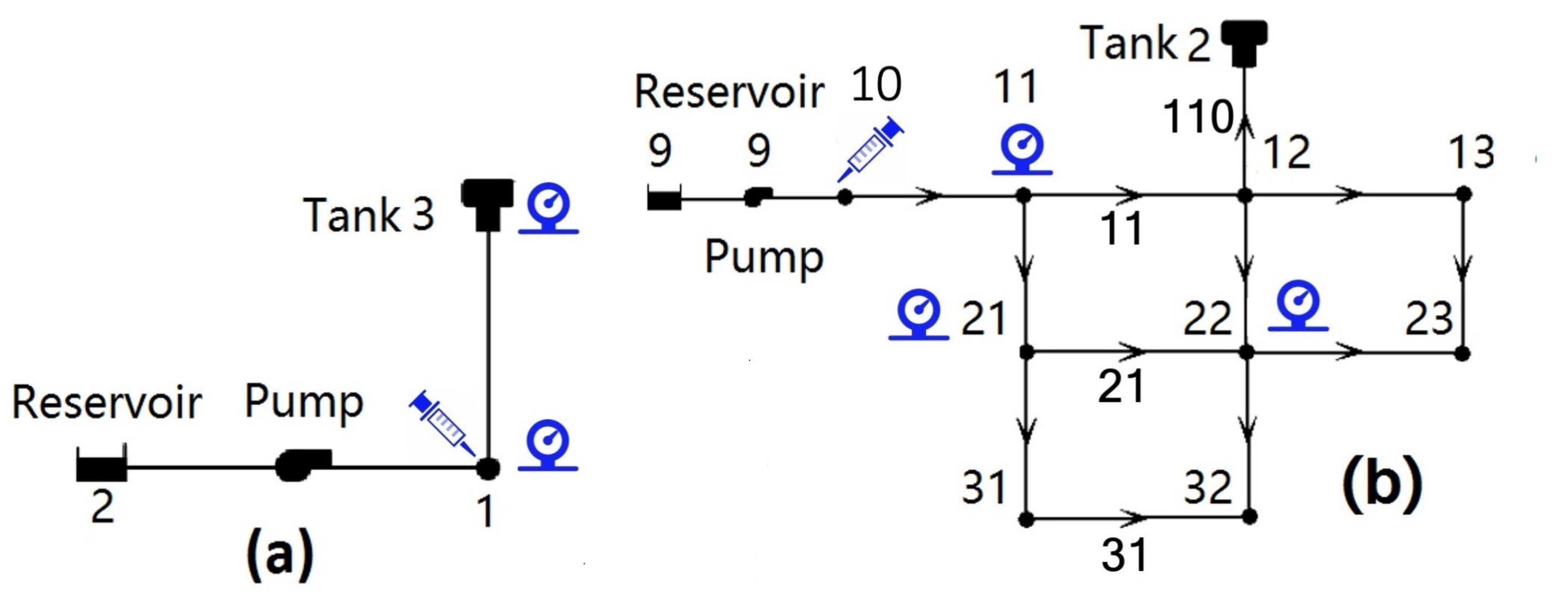}
	\caption{Two tested networks: (a) Three-node network; (b) Net1 network.}
	\label{fig:tested}
\end{figure}

\subsection{Network settings and validation}\label{sec:settings}

\subsubsection{Setup}
The topologies of tested networks are shown in Fig.~\ref{fig:tested}. Three-node network (see Fig.~\ref{fig:tested}a) is a simple, self-designed network for the illustrative purpose that includes one junction J1, one pipe P13, one pump PM21, one tank TK3, and one reservoir R2. A booster station is installed at J1, and two chlorine concentration sensors are put at J1 and TK3 to increase observability, that is $n_u = 1$ and $ n_y = 2$. Similarly, we also tested the Net1 in Fig.~\ref{fig:tested}b which is a looped and more complicated network. Note that Algorithm~\ref{alg:greedy} is used to determine $n_r$ first after setting the energy level goal, and then we select the largest $n_r$ as a common one that works for all SysID algorithms for comparison purposes.

\begin{figure}[t]
	\centering
	\includegraphics[width=0.5\linewidth]{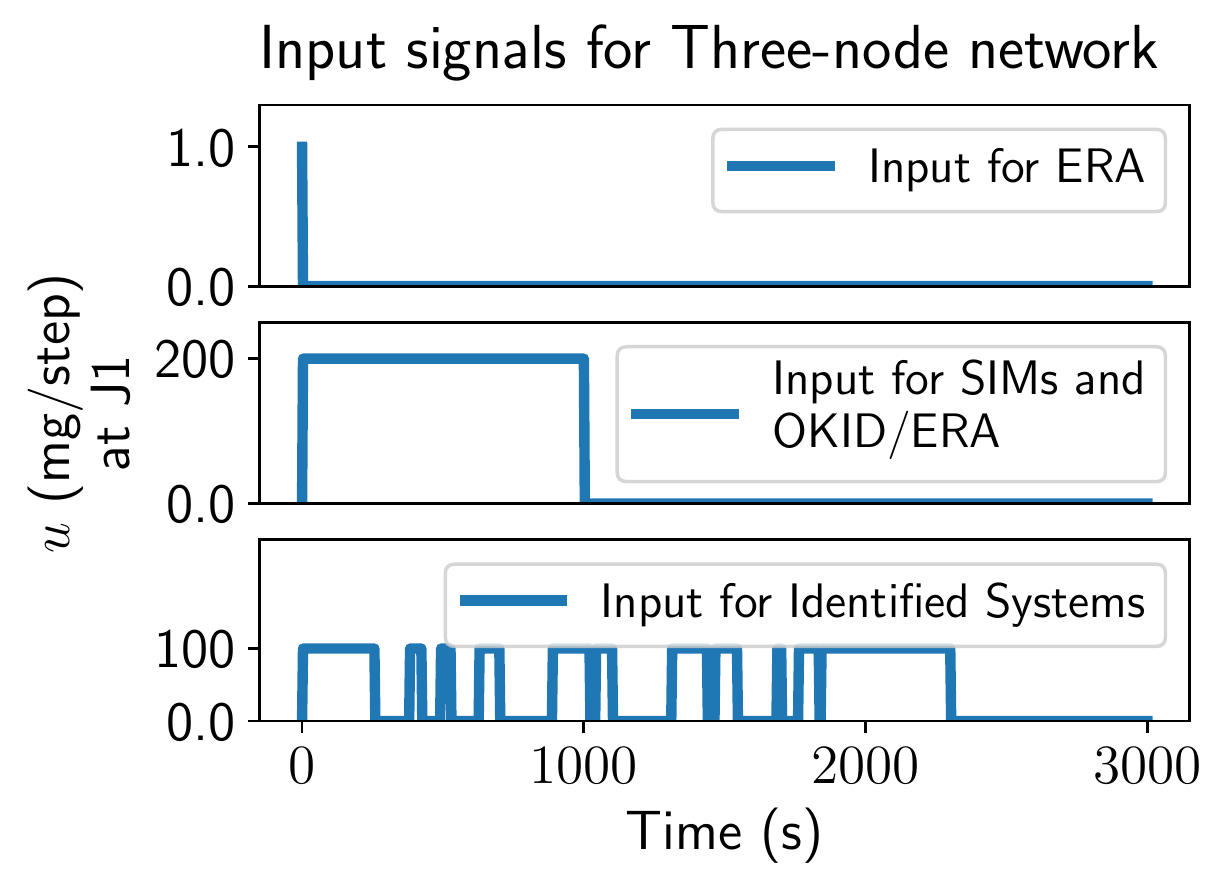}
	\caption{Input signals for tested SysID algorithms (generated by booster station J1 in Fig.~\ref{fig:tested}a). The first two signals are the inputs for SysID algorithms whereas the last one is for validation of identified systems.}
	\label{fig:input}
\end{figure}

Note that we assume the hydraulic time-steps of tested Three-node network and Net1 networks are two hours and six hours which are long enough so that the tested networks can be viewed as an LTI system. Specifically, it takes approximately 10 minutes in the Three-node network and 4 hours and 42 minutes in the Net1 network for input effects to be observed at the outputs.
The system inputs $\m u(k)$, implemented by controlling booster stations, could be an impulse, rectangular, or random signal according to the SysID methods we choose to use in this study; see Fig.~\ref{fig:input}. Moreover, the inputs are divided into test and validation inputs based on the purpose. For a short hydraulic time-step that might cause data completeness issues, we prefer to explore the case in the future even though a possible solution has been proposed in Section~\ref{sec:Challengesolutions}.
Furthermore, zero-initial conditions and zero measurement noises are assumed in this case study.

\begin{table}[t]
	\caption{Basic information of two tested networks.}
	\small
	\centering
	\label{tab:info}
	\setlength\tabcolsep{4pt}
	\renewcommand{\arraystretch}{2}
	\begin{tabular}{c|c|c|c|c}
		\hline
		{\textit{Networks}} & {\textit{\makecell{\# of com-\\ponents$^*$}} } & {\textit{\makecell{Full order\\ $n_x$}}} & {\textit{\makecell{Booster station\\  locations ($n_u$)}}}  & {\textit{\makecell{Sensors\\ location ($n_y$)}}}   \\ \hline
		\textit{\textit{\makecell{Three-node\\network}}} & \makecell{\{1,1,1,\\1,1,0\}}  & 154 & \makecell{J2 \\(1)} & \makecell{J2, TK3 \\ (2)}    \\ \hline
		\textit{Net1} & \makecell{\{9,1,1,\\12,1,0\}}   & 1,293 &  \makecell{J10\\ (1)}  &  \makecell{J11, J21, J22 \\ (3) }   \\ \hline
		\hline
		\multicolumn{5}{l}{\footnotesize{
				\makecell{$^*$Number of each component in WDN: \{$n_\mathrm{J}$, $n_\mathrm{R}$, $n_\mathrm{TK}$, $n_\mathrm{P}$, $n_\mathrm{M}$, $n_\mathrm{V}$\}.} }}
	\end{tabular}%
	\vspace{-0.4cm}
\end{table}

\subsubsection{Validation}
The core idea of the designed experiments in this numerical study is applying the same input signals (for validation purpose) to the original and identified state-space models simultaneously, and then comparing output differences. The accuracy of identified models is validated by the overlapping degree of responses. Mathematically, we use $\mathrm{RMSE} = \sqrt{\frac{1}{m}\sum_{k=1}^{m}{|| \m y_e(k) ||_2^2}}$ to quantify the error, where $ \m y_e(k)$ is defined by $\m y(k)- \hat{\m y}(k)$ is the absolute error, and vectors $\m y(k)$ and $\hat{\m y}(k)$ are measured outputs of the original and identified models.

The original state-space models for the tested two networks available in~\cite{wang2020effective} are built by analysis-based methods and only used to simulate the tested networks and generate output data (i.e., measurement ${\m y}(k)$) when the system input $\m u(k)$ is given. Note that original state-space models are derived when all parameters are perfectly known, and system equations based on the mass and energy balance law are listed. The full order of the original models $n_x$ is high even for the small tested networks; see Table~\ref{tab:info} for details. However, the system order $n_r$ of identified system generated by data-driven based methods is unknown, and it is the only parameter to be determined by users according to the identification procedures introduced in Section~\ref{sec:methods}. We would like to stress that SysID algorithms do not need any other parameters except $n_r$.


\subsection{Performance of SysID algorithms}

\subsubsection{Three-node network}
According to the input of each procedure in Section~\ref{sec:Fundamentals}, the input signals (i.e., $\m u(k)$) for all tested SysID algorithms are shown in Fig.~\ref{fig:input}. In particular, the first subplot is the input for the ERA method that is required to be an impulse signal; inputs for SIMs and OKID/ERA methods can be any kind of signals, but to ensure the data completeness aforementioned, we use a rectangular signal (the second subplot) for the Three-node network. After the system is identified, a random signal shown in the last subplot is a validation input for identified systems.

\begin{figure}[t]
	\centering
	\includegraphics[width=0.55\linewidth]{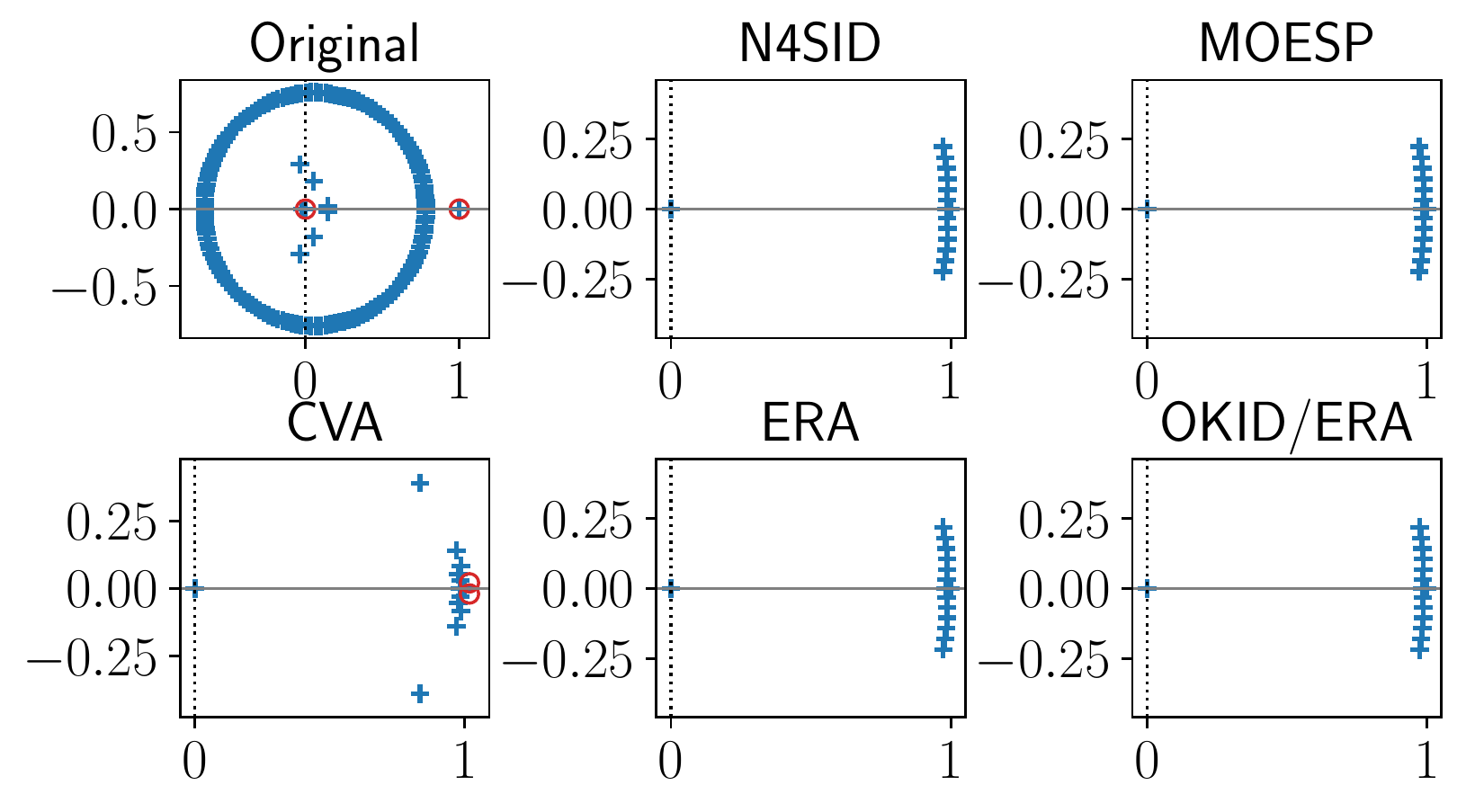}
	\caption{Poles (blue ``+") and zeros (red ``o") in complex plane of original and identified systems via various SysID algorithms. The solid/dashed line is the real/imaginary axis of complex plane.}
	\label{fig:threenodezeropole}
\end{figure}

SIMs and ERA-based methods are tested for this Three-node network, and system order $n_r$ is set to 15 for all SysID algorithms. The model properties of the original and identified systems (e.g., stability, system frequencies, and delay) can be implicitly and roughly compared by a pole-zero plot (see Fig.~\ref{fig:threenodezeropole}). 

Poles represent the behavior of a dynamical system; zeros, on the other hand, stand for how the input signal affects the system itself. The locations of the poles and zeros of the original system in Fig.~\ref{fig:threenodezeropole} indicate that \textit{(i)} the original system is stable since all the 154 poles are inside the unit circle (stability property), \textit{(ii)} the original system has low and high frequencies since the poles appear around the whole unit circle in the complex plane, and \textit{(iii)} huge delay exists in the original system due to several poles are close to the $(0, 0)$ in the complex plane (high-delay property). 
The poles and zeros of the identified systems using SIMs and ERA-based methods are shown in Fig.~\ref{fig:threenodezeropole}. It can be seen that \textit{(i)} the 15 poles of all identified systems are inside the unit circle and stable except the one from the CVA method, \textit{(ii)} all SysID methods keep the low frequencies and high frequencies are neglected, and \textit{(iii)} the high-delay property of the original system is retained by keeping a pole near the $(0, 0)$ in the complex plane. 

\begin{figure}[t]
	\centering
	\includegraphics[width=0.55\linewidth]{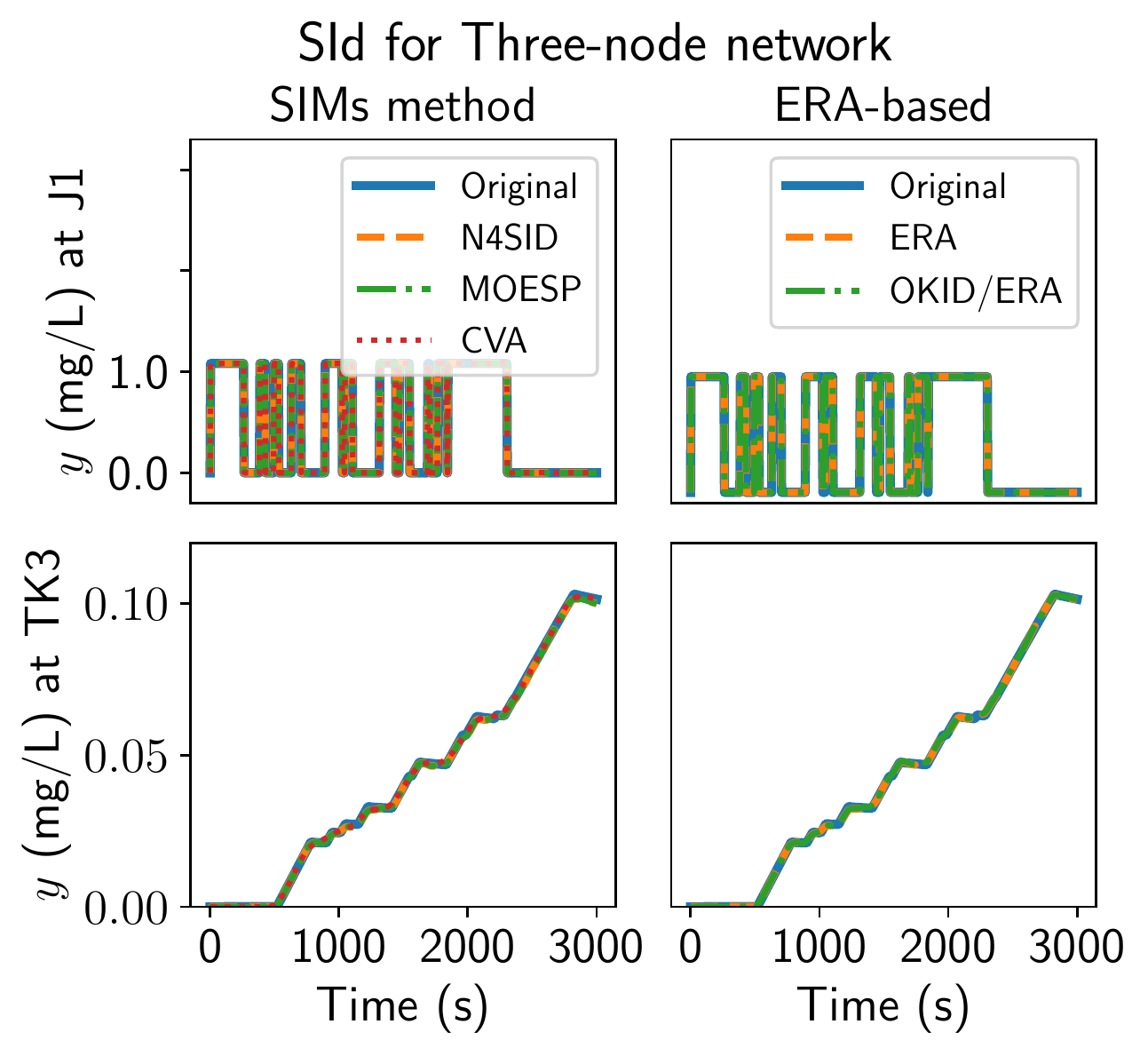}
	\caption{Output comparisons between various SysID algorithms and original system (sensors are installed at J1 and TK3).}
	\label{fig:threenoderesult}
\end{figure}

Another straightforward and accurate way to verify the performance is to compare the actual response to the same input validation signal, that is, simulating the network with chlorine injection. We choose to use the same random signal as a validation input (i.e., the last subplot in Fig.~\ref{fig:input}) for all SysID algorithms (including ERA) such that the results are comparable. To illustrate clearly, the results are grouped by their categories; see Fig.~\ref{fig:threenoderesult}. The left column shows the results from SIMs whereas results from ERA-based methods are in the right column. Observed from the results (sensors are installed at J1 and TK3), we conclude that all methods successfully identified the original state-space system (order $n_x = 154$) but with different inaccuracies. The ERA-based methods are more accurate than the SIMs for the Three-node network. In particular, the $\mathrm{RMSE}$ of N4SID is $0.0712$ which is the smallest while the largest  $\mathrm{RMSE} = 0.323$ is from CVA method; the $\mathrm{RMSEs}$ for other methods are in Table~\ref{tab:RMSE}. 

The corresponding computational time is shown in Table~\ref{tab:my-table}, and the MOESP method takes only 0.21 seconds which is the fastest whereas OKID/ERA takes 3.34 seconds and is the slowest. The computational times indicate the applicability and practicality of using data-driven based methods to identify system models.

\begingroup
\small
\setlength{\tabcolsep}{5pt} 
\begin{table}[t]
	\caption{Computational time (in seconds) of various SysID algorithms for all tested networks.}
	\centering
	\renewcommand{\arraystretch}{1.4}
	\label{tab:my-table}
	\begin{tabular}{c|c|c|c|c|c}
		\hline
		\multirow{2}{*}{\textit{Networks}} & \multicolumn{3}{c|}{\textit{SIMs}} & \multicolumn{2}{c}{\textit{ERA-based method}} \\ \cline{2-6} 
		& \textit{N4SID} & \textit{MOESP} & \textit{CVA} & \textit{ERA} & \textit{OKID/ERA} \\ \hline
		\textit{Three-node network} & 0.21 & 0.27 & 0.51 & 0.78 & 3.34 \\ \hline
		\textit{Net1} & - & - & - & 90.00 & 20.97 \\ \hline \hline
	\end{tabular}%
\end{table}
\endgroup
%
%

\begin{figure}[t]
	\centering
	\includegraphics[width=0.65\linewidth]{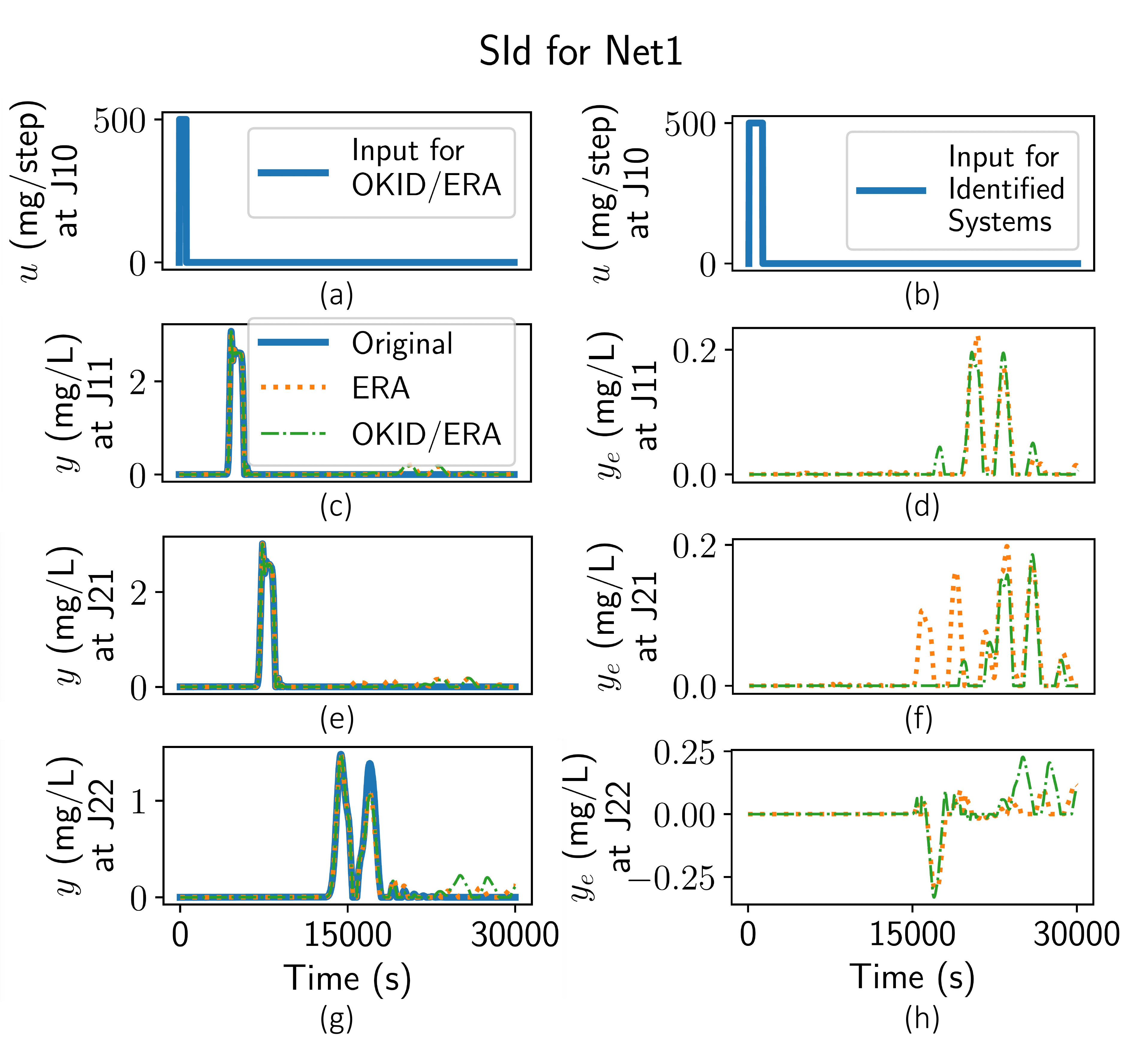}
	\caption{Test and validation inputs for Net1 and output comparisons between SysID algorithms and original system (see booster station and sensors locations in Table.~\ref{tab:info}).}
	\label{fig:net1result1}
\end{figure}

\subsubsection{Net1 network}
We also perform five SysID algorithms to Net1, but all SIMs fail due to the wrong estimation of  Kalman state sequence $\m X_k$ or $\m X_{k+1}$. In the end, they cannot ensure stable identified state-space models (i.e., stable $\m A$ matrices) when solving the least-square problem~\eqref{equ:leastsquare} for large-scale problems. Hence, only results from ERA-based methods (i.e., ERA and OKID/ERA) are presented in Fig.~\ref{fig:net1result1}. Instead of presenting the pole-zero map, we directly show the response to the same input signal. For the ERA method, the test input $\m u(k)$ at J10 is an impulse signal similar to the first figure in Fig.~\ref{fig:input}, and we do not repeat the figure here. For the OKID/ERA method, the test input at J10 is chosen as a rectangle signal lasting for 600 seconds (see the first subplot in Fig.~\ref{fig:net1result1}), whereas the validation input is a different rectangle signal lasting for 1200 seconds (see the second subplot in Fig.~\ref{fig:net1result1}). That is, the booster station injects 80 steps, and each step contains  500 mg of chlorine (1 step lasts for 15 seconds). The injected chlorine parcels travel along the Net1 and reach J11, J21, and J22. The sensors installed at these junctions receive corresponding concentration values as output measurements $y$ for SysID algorithms (the left column of the last three rows in Fig.~\ref{fig:net1result1}). Note that there are two paths to reach J22, and that is why two impulses are observed at J22 (see the last row in Fig.~\ref{fig:net1result1}). Furthermore, the corresponding measurement error $y_e$ is shown in the right column.

From the last three rows in Fig.~\ref{fig:net1result1}, we conclude that ERA and OKID/ERA  successfully identified the water quality models for Net1 since all outputs are overlapping yet with different errors/inaccuracies. The $\mathrm{RMSEs}$ for ERA and OKID/ERA are $3.104$ and $3.402$. Hence, the ERA is more accurate. 
The system order $n_r$ for ERA and OKID/ERA are 115 and 127 which is much smaller than its original full system order $n_x = 1293$. This is due to only the part of the network (controllable and observable part) is identified according to the discussion we mentioned in Section~\ref{sec:discussion}. Furthermore, this water quality model also has the high-delay property, because after the chlorine is injected at J10, and it takes 4680, 7290,  and \{14235, 16860\} seconds (high-delay) to reach all sensors installed at J11, J21, and J22. Note that J22 has two impulses detected; hence, two traveling times values exist.

As for the computational time for Net1, the ERA and OKID/ERA take 90.00 seconds and 20.97 seconds indicating OKID/ERA is much faster than ERA. This is because obtaining Markov parameters in the OKID/ERA procedure is faster. With that in mind, the system identification methods and the corresponding state-space matrices generated from the methods in this paper can still be used to perform water quality regulation or monitoring.

%
%
%
%
\begingroup
\small
\setlength{\tabcolsep}{5pt} 
\begin{table}[t]
		\centering
	\caption{RMSEs of SysID algorithms for Three-node network under different scenarios.}
	\label{tab:RMSE}
	\renewcommand{\arraystretch}{1.4}
	\begin{tabular}{c|c|ccc}
		\hline
		\multirow{2}{*}{\textit{Methods}}    & \multirow{2}{*}{\textit{Algorithms}} & \multicolumn{3}{c}{\textit{Three-node network}}                        \\ \cline{3-5} 
		&                       & \multicolumn{1}{c|}{\textit{Scenario 1}} & \multicolumn{1}{c|}{\textit{Scenario 2}} & \textit{Scenario 3} \\ \hline
		\multirow{3}{*}{\textit{SIMs}} &     \textit{N4SID}                   & \multicolumn{1}{c|}{0.0332}   & \multicolumn{1}{c|}{0.0712}   & 0.0107   \\ \cline{2-5} 
		&        \textit{MOESP}               & \multicolumn{1}{c|}{0.0332}   & \multicolumn{1}{c|}{0.0516}   &   0.0105  \\ \cline{2-5} 
		&          \textit{CVA}             & \multicolumn{1}{c|}{0.0183}   & \multicolumn{1}{c|}{0.323}   & ---   \\ \hline
		\multirow{2}{*}{\textit{\makecell{ERA-based\\method}}}       &      \textit{ERA}                  & \multicolumn{1}{c|}{0.0224}   & \multicolumn{1}{c|}{0.0152}   & 0.0037   \\ \cline{2-5} 
		  & \textit{OKID/ERA}         & \multicolumn{1}{c|}{0.0234}   & \multicolumn{1}{c|}{0.0154 }   &  0.0037   \\ \hline \hline
	\end{tabular}
\end{table}
\endgroup

\subsection{Possible factors affecting SysID algorithms}
From the procedures, we know that there are two main factors determined by users that might have an impact on the final SysID results: system inputs and system order $n_r$. Hence, we use Three-node network as an example to illustrate the impacts. To this end, three scenarios are designed as
\begin{itemize}
	\item Scenario 1: $\m u = $ rectangular signal; $n_r = 15$
	\item Scenario 2: $\m u = $ random signal; $n_r = 15$
	\item Scenario 3: $\m u = $ rectangular signal; $n_r = 40$.
\end{itemize}
Note that the input signal is not changed for ERA for all three scenarios since it always requires an impulse signal as input. 

The final results are shown in Table.~\ref{tab:RMSE}. Comparing Scenarios 1 and 2, we notice that the results are similar indicating the input signal does not impact the results significantly. Comparing Scenarios 2 and 3, all $\mathrm{RMSEs}$ are apparently reduced which indicates improving the system order $n_r$ helps to improve the accuracy of SysIDs. However, we also would like to emphasize that the larger $n_r$ causes stability problems for SIMs, that is the SIMs tend to be unstable after $n_r$ is increased to a certain number (see CVA in Table.~\ref{tab:RMSE} fails to give a stable $\m A$ matrix). Even though the phenomena are not observed for the ERA-based methods, we still prefer to keep the $n_r$ as small as possible.


\section{Conclusions and Future Work}~\label{sec:limitations}
Given the thorough  tests in the previous section, the following observations are made, thereby answering the posed research questions in Section~\ref{sec:casestudy}:

\begin{itemize}
	\item[-] {A1:}  Water system operators are able to obtain water quality models of a WDN only utilizing the collected input-output data. 
	\item[-] {A2:}  In general, ERA-based methods are more accurate and stable than SIMs (especially for high-order systems).   
	\item[-] {A3:} After obtaining the system model, the water system operator can potentially use them to simulate the water quality directly and perform quality control easily using popular control algorithms such as model predictive control.
\end{itemize}

In summary, this paper explores the applicability of classical system identification methods in water quality dynamics for the first time. Furthermore, the properties of the water quality models are studied and proposed, and the ensuing challenges caused by these properties and the corresponding solutions are discussed. The numerical case study shows the performance of each tested SysID algorithm and the possible factors affecting the results. To make SysID algorithms applicable, a constant chlorine decay rate in WDNs and the large-enough hydraulic time-steps of tested networks are assumed so that the tested networks can be viewed as an LTI system. Moreover, zero-initial conditions and zero measurement noises are assumed to meet the requirements of SysID algorithms. Future work will focus on solving the DT-LTV system identification for water quality dynamics considering significant measurement noise, uncertainties, leaks, and multi-species, giving sensor placement strategy when solving for high-delay issue, and combining with MPC algorithm to solve water management issues. 

\section{Supplemental Materials}
The detailed system identification procedures for water quality models including our proposed binary search method to determine system order are available online in the ASCE Library (\href{http://ascelibrary.org/}{ascelibrary.org}).

\section*{Data Availability Statement}

Some or all data, models, or code that support the findings of this study are available from the corresponding author upon reasonable request (e.g., the tested Three-node and Net1 networks and their discrete-time state-space water quality models, the code of SIM, ERA, and OKID-ERA algorithms).

%

	\bibliographystyle{IEEEtran}
	\bibliography{mybib}

\end{document}